\documentclass{article}
\usepackage[dvipdfmx]{graphicx}
\usepackage{spconf,amsmath,amssymb,url}
\usepackage{tikz}
\usepackage{comment}
\usepackage[
    backend=biber,
    style=ieee,
    maxbibnames=3,
    maxcitenames=3,
    doi=false,isbn=false,url=false,eprint=false
]{biblatex} 
\usepackage{multirow}
\usepackage[subrefformat=parens]{subcaption}

\usepackage{hyperref}
\hypersetup{
    colorlinks = false,
}

\DeclareSourcemap{
	\maps[datatype=bibtex, overwrite=true]{
		\map{
			\step[fieldsource=booktitle,
			match=\regexp{.*Interspeech.*},
			replace={Proc. INTERSPEECH}]
			\step[fieldsource=journal,
			match=\regexp{.*INTERSPEECH.*},
			replace={Proc. INTERSPEECH}]
			\step[fieldsource=booktitle,
			match=\regexp{.*ICASSP.*},
			replace={Proc. ICASSP}]
			\step[fieldsource=booktitle,
			match=\regexp{.*icassp_inpress.*},
			replace={ICASSP (in press)}]
			\step[fieldsource=booktitle,
			match=\regexp{.*International.*Conference.*on.*Acoustics,.*Speech.*and.*Signal.*Processing.*},
			replace={ICASSP}]
			\step[fieldsource=booktitle,
			match=\regexp{.*International.*Conference.*on.*Learning.*Representations.*},
			replace={ICLR}]
			\step[fieldsource=booktitle,
			match=\regexp{.*International.*Conference.*on.*Machine.*Learning.*},
			replace={Proc. ICML}]
			\step[fieldsource=booktitle,
			match=\regexp{.*Automatic.*Speech.*Recognition.*and.*Understanding.*},
			replace={Proc. ASRU}]
			\step[fieldsource=booktitle,
			match=\regexp{.*Spoken.*Language.*Technology.*},
			replace={Proc. SLT}]
			\step[fieldsource=booktitle,
			match=\regexp{.*Speech.*Synthesis.*Workshop.*},
			replace={Proc. SSW}]
			\step[fieldsource=booktitle,
			match=\regexp{.*workshop.*on.*speech.*synthesis.*},
			replace={Proc. SSW}]
			\step[fieldsource=booktitle,
			match=\regexp{.*Advances.*in.*neural.*information.*processing.*},
			replace={Proc. NIPS}]
			\step[fieldsource=booktitle,
			match=\regexp{.*Advances.*in.*Neural.*Information.*Processing.*},
			replace={Proc. NIPS}]
			\step[fieldsource=booktitle,
			match=\regexp{.*AAAI.*},
			replace={Proc. AAAI}]
			\step[fieldsource=booktitle,
			match=\regexp{.*Workshop.*on.* Applications.* of.* Signal.*Processing.*to.*Audio.*and.*Acoustics.*},
			replace={Proc. WASPAA}]
			\step[fieldsource=booktitle,
			match=\regexp{.*International.*Conference.*on.*Language.*Resources.*and.*Evaluation.*},
			replace={Proc. LREC}]
			\step[fieldsource=journal,
			match=\regexp{.*Spontaneous.*Speech.*Processing.*and.*Recognition},
			replace={Proc. SSPR}]
			\step[fieldsource=publisher,
			match=\regexp{.+},
			replace={{}}]
			\step[fieldsource=month,
			match=\regexp{.+},
			replace={{}}]
			\step[fieldsource=location,
			match=\regexp{.+},
			replace={{}}]
			\step[fieldsource=address,
			match=\regexp{.+},
			replace={{}}]
			\step[fieldsource=organization,
			match=\regexp{.+},
			replace={{}}]
			\step[fieldsource=doi,
			match=\regexp{.+},
			replace={{}}]
			\step[fieldsource=url,
			match=\regexp{.+},
			replace={{}}]
			\step[fieldsource=editor,
			match=\regexp{.+},
			replace={{}}]
		}
	}
} 

\title{Text-to-speech synthesis from dark data with\\
evaluation-in-the-loop data selection}
\name{Kentaro Seki, Shinnosuke Takamichi, Takaaki Saeki, and Hiroshi Saruwatari\thanks{This work is supported by JSPS KAKENHI 22H03639 and Moonshot R\&D Grant Number JPMJPS2011. We also appreciate Kenta Udagawa of the University of Tokyo for his help.}}
\address{The University of Tokyo, Japan.}

\bibliography{references}

\makeatletter
\newcommand{\figcaption}[1]{\def\@captype{figure}\caption{#1}}
\newcommand{\tblcaption}[1]{\def\@captype{table}\caption{#1}}
\makeatother

\begin{document}
\ninept
\maketitle
\setlength{\tabcolsep}{1mm} 

% abstract
\begin{abstract} \vspace{-1mm}
This paper proposes a method for selecting training data for text-to-speech (TTS) synthesis 
  from dark data.
TTS models are typically trained on high-quality speech corpora that cost much time and money for data collection,
  which makes it very challenging to increase speaker variation.
In contrast, there is a large amount of data whose availability is unknown (a.k.a, ``dark data''),
  such as YouTube videos.
% 我々が話者バリエーションを増やすためにすべきことは、コーパスの拡張ではなくダークデータの活用方法の模索である。
% To increase speaker variation,
%  we should explore the utilization of such data rather than expanding TTS corpora.
% TTSコーパス以外の音声を活用する方法として、自動音声認識コーパスから音響品質に基づいてデータ選択を行う方法が提案されている。
To utilize data other than TTS corpora, 
  previous studies have selected speech data from the corpora on the basis of acoustic quality.
% 提案のモチベーション
% 1. However, given the robustness of recent TTS models to data noise, we should perform the data selection on the importance of the training data to the given TTS model, not the quality of speech itself.
% 2. Audio qualityそのものではなく，そのtraining dataとしての重要度 (synthetic speech qualityへの寄与度) に基づいてdata selectionを行う方がより直接的であり，質の高いTTS training dataの構築を行えると考えられる
% -> しかし、近年はデータノイズに頑健なTTSモデルが提案されており、音声の質そのものではなく、与えられたTTSモデルに対する学習データとして重要度でデータ選択を行うべきである。
However, considering that TTS models robust to data noise have been proposed, 
  we should select data on the basis of its importance as training data to the given TTS model, not the quality of speech itself.
% ↓論文の手法を簡潔に述べる文
% 1.The perceptual quality of synthesized speech is automatically predicted, and the TTS corpus is determined based on the quality of the synthesized speech under the given TTS model. 
% 2.We leverage an automatic quality asessment model for synthetic speech and develop a regression model to perform both speaker-wise and utterance-wise data selection.
% -> 合成音声の品質を予測する回帰モデルを構築し発話レベルのデータ選別を行うことで、TTSコーパスを構築する。
% 省いた要素：モデルごとに異なるコーパスを構成すること
Our method with a loop of training and evaluation 
  selects training data on the basis of the automatically predicted quality of synthetic speech of a given TTS model.
Results of evaluations using YouTube data reveal that our method outperforms the conventional acoustic-quality-based method.

\end{abstract} \vspace{-1.2mm}

% keyword

% Section 1: Introduction
\vspace{-3mm}
\section{Introduction}
\begin{figure}[t]
  \centering
    \vspace{-0.5mm}
    \includegraphics[width=0.6\linewidth]{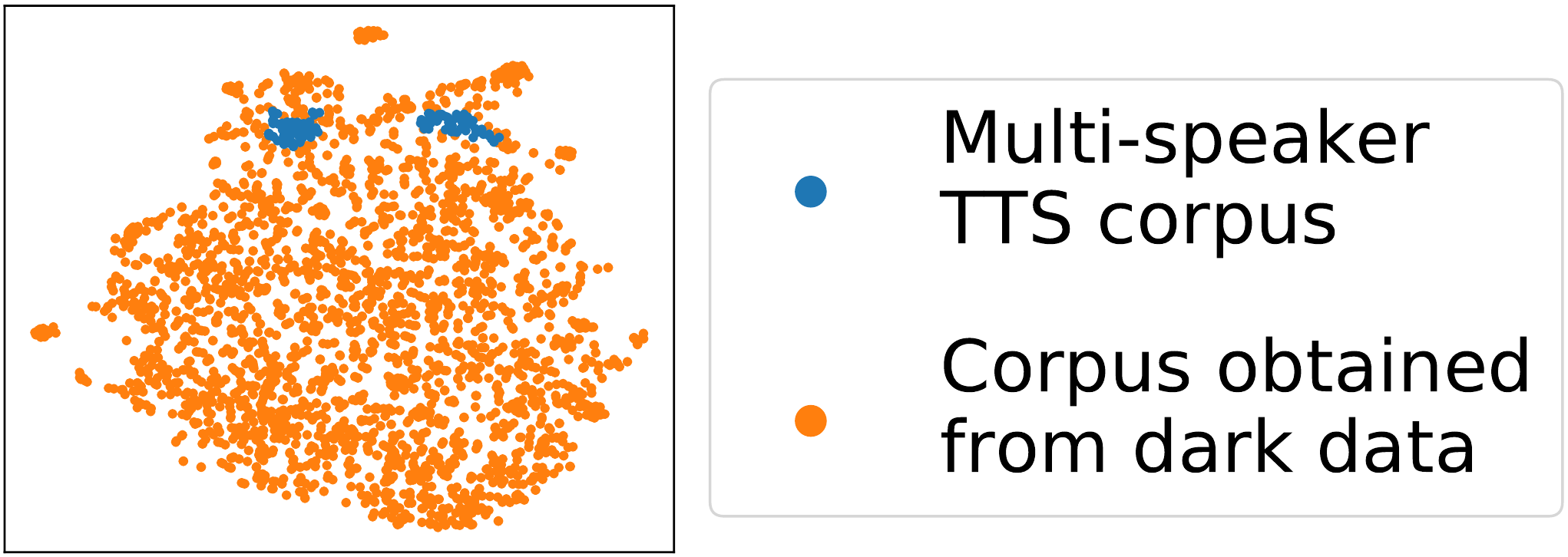} 
    \vspace{-2.8mm}
    \caption{ 
      Distributions in speaker space for multi-speaker TTS corpus~\cite{JVS} and dark data~(e.g., data from YouTube~\cite{JTubeSpeech}).
      Compared with speakers in the massive dark data, speakers in TTS corpus are very limited and localized.}
    \label{fig:話者分布}
    \vspace{-5.5mm}
\end{figure}

\vspace{-2.5mm}
With large speech corpora and the development of sequence-to-sequence models, 
  recent text-to-speech~(TTS) models have achieved human-like speech synthesis 
  at a level comparable to human speech utterances~\cite{Tacotron2, TransformerベースでTaco2並の性能,%microsoftのTTSで自然音声に近いMOS値,DiffusionベースでTaco2並,
  %VAE等を入れて人間音声に匹敵するレベルの合成,
  NaturalSpeech}.
Multi-speaker TTS %an extension of the TTS framework, 
  can synthesize speech with the desired speaker characteristics 
  by inputting speaker information in addition to text~\cite{
    %DeepVoice2,
    DeepVoice3, 
    %MultiSpeakerTTSwithDVector,
    話者認証からMultiSpeakerへの転移,
    %Multispeech,%moss2020boffin,
    cooperZeroShotMultiSpeakerEmbedding,
  }.
TTS training typically requires pairs of text and speech. 
The speech data is recorded in a well-designed environment~\cite{veaux2016superseded,JSUT,JVS}, e.g., a recording studio with a professional speaker. 
Furthermore, constructing a multi-speaker TTS corpus is even more burdensome. 
This fact significantly limits the variety of speakers that TTS can synthesize. 
Fig.~\ref{fig:話者分布} is an example of 
  speaker distributions
  of a famous multi-speaker TTS corpus (JVS~\cite{JVS}) and one we deal with in this paper. 
% これらの分布は自然音声のxvectorをt-SNEで次元削減して得られたものである。
This is a t-SNE~\cite{t-SNE} plot of $x$-vectors~\cite{xvector}
  of natural speech in the two corpora.
Typical multi-speaker TTS corpora only cover limited speaker variation.

In contrast, there is a large amount of speech data (e.g., YouTube videos) with unknown potential for machine learning, known as \textit{dark data}~\cite{
    trajanov2018dark,
    schembera2020dark,
  }, on the Internet. 
We expect to solve the small speaker variation problem if we can fully automate TTS model training from dark data. Related to this, methods have been proposed for building TTS corpora from datasets other than TTS corpora~\cite{
    LibriTTS,
    bakhturina2021hi,
    LibriVoxからドイツ語,
    TEDから英語コーパス,
}. 
% These methods quantify the sound quality of each speech in the dataset and determine a TTS corpus based on the threshold of the sound quality. 
These methods determine a TTS corpus on the basis of the acoustic quality of each utterance. 
On the other hand, recent TTS models are becoming more robust to data noise than the old-fashioned parametric models~\cite{DenoiSpeech, DRSpeech, DenoiSpeech的なやつ}. In other words, low-quality speech data does not necessarily have a negative impact on TTS model training. In light of the above, TTS corpus construction should be based on the importance of the speech data as training data for a given TTS model, not on the quality of speech data itself.

This paper proposes an evaluation-in-the-loop data selection method for TTS model training from dark data.
% 提案手法は、与えられたTTSモデルから出力される合成音声の知覚的品質を用い、学習データ候補のそれぞれを音質ではなく、学習データの重要度としてスコア化するものである。
Our method uses a pseudo score on the perceptual quality of synthetic speech output from a given TTS model and scores each of the training data candidates in terms of the importance of training data rather than acoustic quality. 
% 回帰モデルを用いたスコアリングの話
The training data candidates are filtered by the importance.
% 元の文：Scoring with a regression model enables utterance-wise filtering.
% この手順では、自然さに関する平均的な意見スコアを自動的に予測する事前学習済みモデルを用いて、選択と学習のループを自動化する。
The procedure automates the selection and training loop using a pre-trained model that automatically predicts 
  the mean opinion score~(MOS) of synthetic speech on naturalness. 
% また、ダークデータの収集からTTSモデル学習までの完全自動化を実現し、TTS利用のためのダークデータの選別方法を提案する。
This paper also proposes a method for pre-screening dark data for TTS use, achieving a fully automated process from dark data collection to TTS model training. 
% ダークデータを用いた実験により、提案手法が従来のデータ品質に基づく手法よりも良好な結果を得ることができた。
We conduct experiments with dark data, in particular, actual data obtained from YouTube.
The results show that our evaluation-in-the-loop method achieves better than the conventional acoustic-quality-based method. 
% 現実世界がデカすぎることを示す図
% 紙面の都合で段落を買えない
% 以下で我々の貢献を述べる
The contributions of this work are as follows:
\begin{itemize} \vspace{-1.2mm} \itemsep -0mm \leftskip -5mm
    \item We propose an evaluation-in-the-loop data selection method for TTS model training,
      that enables more efficient selection than the conventional acoustic-quality-based method.
    % 回帰モデルを用いたフィルタリングにより、話者レベルではなく発話レベルでのデータ選択が実現する。
    % \item Our method enables data selection at utterance-wise, not at speaker-wise.
    \item We conduct experiments using actual data downloaded from YouTube 
      and demonstrate the validity of the proposed method.
\end{itemize} \vspace{-1mm} 

% Section 2: Related Work
\vspace{-3.3mm}
\section{Related work} 
\label{sec:related_work}
\vspace{-2.1mm}
\textbf{Multi-speaker and noise-robust TTS.}
% Multi Speaker TTS は、話者性を制御するために
%   話者表現を用いる
Multi-speaker TTS uses speaker representations to control the speaker of the synthetic speech~\cite{DeepVoice3},
  e.g., the $x$-vector~\cite{xvector} of automatic speaker verification (ASV); 
  we follow this method in this paper. 
Also, a few-shot speaker adaptation, voice building of a new speaker from a small amount of speech data~\cite{cooperZeroShotMultiSpeakerEmbedding}, 
  is a promising application of multi-speaker TTS. 
To increase the number of speakers by using diverse training data, 
  there exist methods for noise-aware TTS training~\cite{DenoiSpeech,DRSpeech,DenoiSpeech的なやつ}.

\textbf{Building TTS corpus on automatic speech recognition~(ASR) corpus.}
The use of more speakers' data in training increases 
  the speaker variation in TTS~\cite{話者認証からMultiSpeakerへの転移}. 
Methods based on selecting data from ASR corpora 
  have been proposed. 
The successful examples are based on acoustic noise~\cite{話者認証からMultiSpeakerへの転移} and text label noise (i.e., a mismatch between text and speech)~\cite{LibriTTS}. 
% These methodologies help construct a TTS corpus from an ASR corpus. 
However, they are based only on data quality and are performed independently from TTS models. 
They are not appropriate considering the noise robustness of the models, which we described above.

In addition, dark data has the potential to build massive corpora almost fully automatically, and indeed, successful examples of this have been reported in the construction of ASR corpora~\cite{galvez2021peoplespeech,chen2021gigaspeech}. 

\textbf{Prediction of perceptual quality in synthetic speech.}
Methods for automatically predicting the subjective evaluation score 
  of synthetic speech have been studied 
  to reduce the evaluation cost for TTS~\cite{moller10mosprediction,patton16automos}. 
The challenge in automatic prediction is generalization performance, i.e., 
  the robustness of the prediction against divergence between training and evaluation data. 
The recent deep learning-based models (used in this paper) have 
  relatively good accuracy in predicting the relative rank between speech samples, 
  even if the prediction value itself is not accurate~\cite{huang22voicemos}.

% Section 3: Proposed Methods
\vspace{-4.5mm}
\section{proposed method} \vspace{-1.mm}

\begin{figure}[t]
% \begin{center}
\centering
\includegraphics[width=0.9\linewidth]{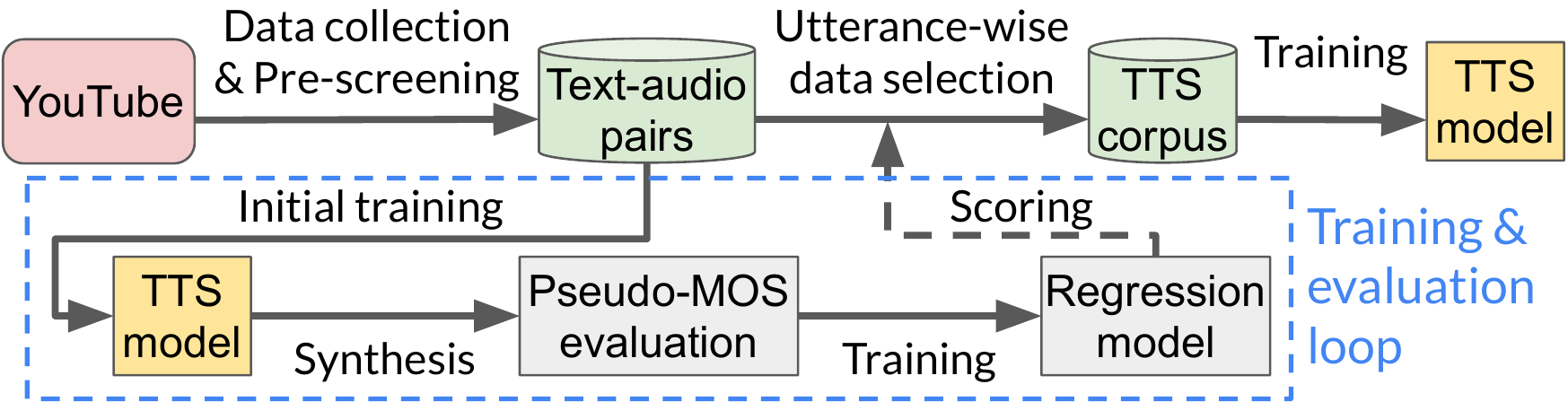}
\vspace{-2.5mm}
\caption{%\centering
  % 提案手法の手順
  Procedure of proposed method. 
  % YouTubeからdark dataを取得し、
  %   TTSモデルの学習と合成音声の評価を通して各発話を評価する。
  We obtain dark data from YouTube
    and evaluate each utterance through loop
      of TTS model training and synthetic-speech evaluation.
  % 発話レベルのスコアリングによってコーパスを構築する。
  It finally builds TTS corpus from dark data by utterance-wise filtering.
}
\vspace{-4mm}
\label{fig:フロー図}
% \end{center}
\end{figure}

% Fig.~\ref{fig:フロー図} shows an overview of the proposed method. 

\vspace{-2mm}
\subsection{Data collection and pre-screening} \label{sec:prescreening} \vspace{-2mm}
% 3.1. データ取得
% 最初にやることはスピーチデータの取得とpre-screeningである。
As shown in Fig.~\ref{fig:フロー図},
  the first step is to collect text-audio pairs 
  and pre-screen dark data to filter too low-quality data for TTS training.
% youtubeから取ってきて、ASRとASVの方法を結合させた。
We collect a dataset from YouTube 
  and combine data screening methods for ASR and ASV.
% 以下でそれぞれの方法を簡潔に述べるが、詳細はThe paperを見てくれ。
We briefly describe each method below,
  but see the paper~\cite{JTubeSpeech} for more detail.

\textbf{Cleansing based on text-audio alignment accuracy.}
% 音声合成の学習データにはテキストラベルが必要である。
Speech data in TTS training data 
  must align well with its transcription.
We calculate the connectionist temporal classification~(CTC) score~\cite{CTC-segmentation} 
  to quantify how well speech utterances fit 
    to text (YouTube subtitles in this case). 
Audio data is divided into each utterance by CTC segmentation,
  and utterances with lower scores are eliminated.

% \textbf{Speaker compactness.}
\textbf{Cleansing based on speaker compactness.}
% 各話者について複数の発話があることが望ましい
Having multiple utterances for each speaker is desirable for TTS training data. 
% そこで各発話グループごとにdベクトルの分散を計算し、
%   複数の発話において発話表現がどの程度安定しているかを定量的に評価する。
We calculate the $x$-vector~\cite{xvector} variance within each utterance group (utterances belonging to one video in the YouTube case) to quantify how stable the speaker representation is in the multiple utterances.
% ここで音声は予めutteranceレベルに分割されている。
% Here, audio data is divided into each utterance by CTC segmentation.
% スコアの低い発話はフィルタリングされる。
Utterance groups with lower this scores are filtered out as well as above.
% フィルタリング後の発話は、一つのグループ内の発話は固有の話者によるものとみなされる。
After the filtering, 
  utterances within one group are considered to be by a unique speaker.
% なお、この方法では同じ話者
Note that this method implicitly rejects single speaker utterances
    with multiple styles or large fluctuations in style 
    because the $x$-vector varies greatly depending on the speaking style~\cite{williams2019disentangling}, 
      even for the same speaker. 
% 今回はスタイルではなく話者に着目するためこのようなことをした
We used this method because we focus on speaker variation 
  rather than speaking-style variation.

\vspace{-3.2mm}
\subsection{Data selection using automated evaluation-in-the-loop} 
\vspace{-2mm}
Our method aims to estimate the quality of each data in terms of \textit{training data for the given TTS model}.
% 品質スコアは、音声データを用いて学習したTTSモデルの合成音声の擬似的な知覚的自然さとして予測される。
% TTSモデルの学習と合成音声の自動評価の繰り返しと見ることができる。
For each speech data,
  we predict the pseudo perceptual naturalness of the synthetic speech 
    when the TTS model is trained on the speech data.
We use this score for the quality score of each data.
The TTS model is finally trained using the TTS training data obtained via the evaluation-in-the-loop selection.

\vspace{-3.5mm}
\subsubsection{Initial training using pre-screened data} 
\vspace{-2.5mm}
% seq-to-seqのMulti speaker TTSモデルを学習した。
% 学習にはprescreeningで得られたデータすべてを用いた。
We perform the initial training of the given TTS model, using all the data obtained from the pre-screening. 
% === 高道 ===
% We perform the initial training of the given TTS model, using all the data obtained from the pre-screening. 

\vspace{-4mm}
\subsubsection{Evaluating synthetic speech quality} \vspace{-2mm}
\label{section:回帰モデルに関する説明}
% 学習したモデルの合成音声の品質を評価する。
We evaluate the quality of speech synthesized by the initially-trained TTS model.
% 単純な方法はloss functionを見ることである。
A simple evaluation method is to calculate the value of the loss function for each utterance during training (e.g., the distance between the ground-truth and the predicted features).
% しかし、この距離は必ずしも品質とは対応しない。
However, 
  the distance does not necessarily correspond to the perceptual quality of the synthesized speech~\cite{hayashi2021espnet2,weiss2021wave}.
% 不要だと思ったのでコメントアウト
% さらに、この方法はover fittingに弱い。
% Furthermore, this method is also prone to over-fitting; it is inappropriate for the purpose. 

% もう一つの方法は主観評価をすることである
Since MOS directly reflects perceptual quality, 
  another method is to conduct subjective evaluation.
However, it does not scale well for training with a massive amount of data.
We leverage pseudo MOS predicted with an automatic MOS prediction model. 
% section2で述べた通り、最近のMOS予測モデルの汎化性能の高さから、ここでの利用が可能だと考えられる。
As described in Section~\ref{sec:related_work}, current MOS prediction models have achieved a generalization ability. 
Therefore, we use a pseudo MOS score of subjective evaluation on naturalness
  predicted by a pre-trained MOS prediction model.

% この評価結果は、
%   次のステップで学習データと合成音声の品質の関係を推定するために使う。
This evaluation is used to 
  determine the score of each training data,
  as described in the following Section~\ref{sec:発話レベルスコアリング}.
% すなわち、データそれぞれが合成音声の品質に与える影響の違いを見たいのである。
In other words, 
  this evaluation aims to estimate the difference in 
  the effect of each data on the quality of the synthesized speech.
% 単純な方法は各データと同じ音声を合成させる方法である
The simplest method is to synthesize training data sentences 
  and to filter out sentences with lower scores.
% 文ごとのを使わなかった理由：
However, this method uses different sentences among speakers.
It is inappropriate because
   1) the pseudo MOS score changes depending on 
         the sentence to be synthesized~\cite{UTMOS} and
   2) a sentence set greatly varies among speakers in dark data.
% ここで、我々は話者ごとに品質評価をする
Therefore,
  we evaluate the quality of synthesized speech for each speaker,
  using common sentences.
% 発話内容と独立にデータ品質による合成品質の違いを知ることができる。
In this way, it is possible to quantify 
  the difference in synthetic speech quality of each speaker, 
  without depending on the speakers' utterances.
% 具体的には
%   学習データに含まれない複数の文の読み上げ音声を評価した。
% この値は話者ごとに平均され、話者ごとスコアとして使われた。
Therefore,
  we evaluate the quality of synthesized speech for each speaker
  using common sentences not included in the training data,
  and averages of the values are used for each speaker.

% "In this way" 〜に統合
% Specifically, we obtain pseudo MOS scores using speech synthesized from each of the common sentences and average the scores within each speaker. This methodology makes it possible to avoid the bias problem and obtain unbiased speaker-wise pseudo MOS.

\vspace{-3.5mm}
\subsubsection{Quantifying score as training data for each utterance}
\vspace{-2mm}
\label{sec:発話レベルスコアリング}
% グラフ
\begin{figure}[t]
%\begin{center}
\centering
\includegraphics[width=1.0\linewidth]{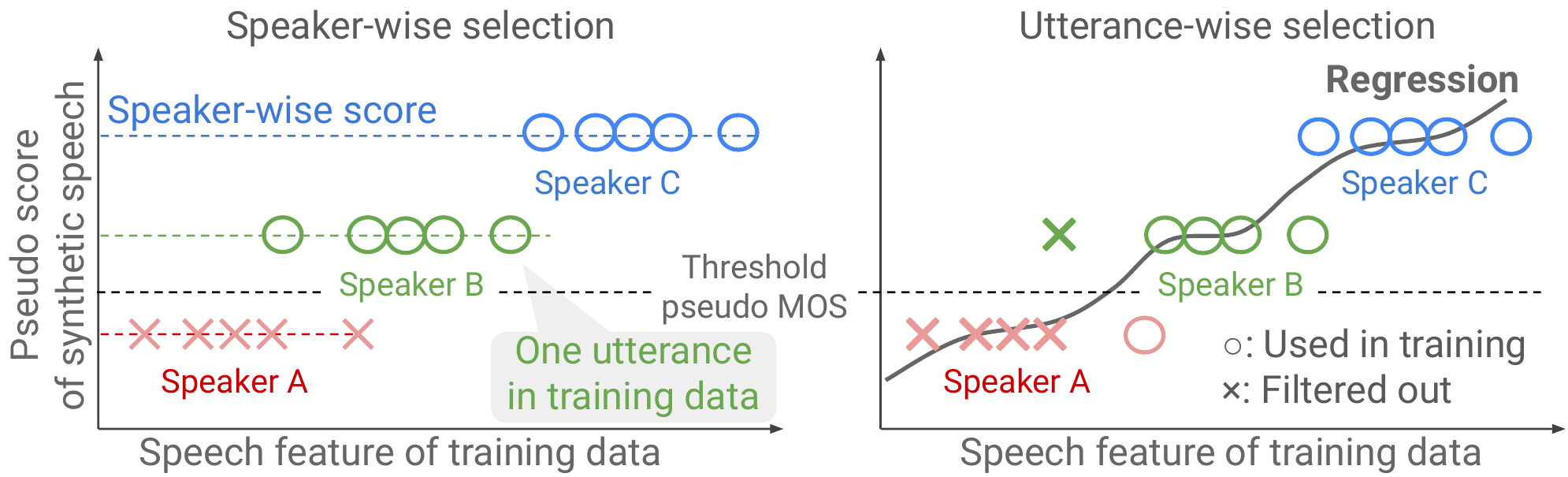}
 \vspace{-6.5mm}
 \caption{
   % 話者レベル vs 発話レベル
   Comparison of speaker-wise and utterance-wise selection.
   % 音響品質から評価するため、合成音声の品質が低い話者であっても音響品質の高い発話は持ってこれる。嬉しい
   With regression,
     we filter out low-score utterances 
     even if speaker's pseudo MOS is high.
  }
 \vspace{-4mm}
 \label{fig:回帰モデルの説明}
%\end{center}
\end{figure}

% 話者ごとの疑似スコアを使ってフィルタリングする。
We filter the training data on the basis of the obtained speaker-wise pseudo MOS.
% 最も単純な方法は話者レベル評価値をそのまま値として使うこと
The simplest method is to filter at the speaker level on the basis of the values,
  i.e., filtering out speakers with lower pseudo MOSs. 
% だが、同一話者でも品質に差があるし、フィルタリングは発話レベルで行うべき。
However, since the data quality varies within the same speaker, 
  filtering should be performed at the utterance level. 
% すなわち低品質話者からであっても高品質なデータがあったら選ぶべきだ
In other words, 
  we should filter out low-quality utterances even if speaker's pseudo MOS is high,
  and vice versa as shown in Fig.~\ref{fig:回帰モデルの説明} .

% この目的のために、データから話者ごと疑似MOSを予測するモデルを組み立てる。
For this purpose, we train a regression model that 
  predicts the speaker-wise pseudo MOS from each utterance in the training data. 
% 我々は音響的に似たデータは自然性に同じような貢献をすると考えられる。
We assume that acoustically similar training data will achieve similar naturalness in the synthesized speech. 
% 回帰モデルは音響的に似たデータに対して近い値を予測することが期待されるため、
Furthermore, we assume that a regression model predicts close values for acoustically similar data.
% これらの動機から、我々は回帰モデルを使ってデータを評価する。
These assumptions motivate us to use a regression model.

% この回帰モデルをpre-screened data全てで学習
We train this regression model on all the pre-screened data.
% 回帰モデルを用いることの効果は実験によって確認する。
We confirm the effect of using a regression model 
  in the experiment described in Section~\ref{sec:PseudoMOSExperimentResult}.

\vspace{-3.5mm}
\subsection{Training data selection and re-training} \vspace{-2mm}
% 発話レベルフィルタリング
We evaluate the training data at the utterance level with the regression model.
% この回帰モデルによって、閾値を下回る発話をfilter outする。
Then, utterances with lower scores are filtered out.
% その後選択されたデータによってTTSモデルの学習をやり直した。
Finally, we retrain the TTS model with the filtered data.
% === 高道 ===
% Using the trained regression model, we calculate scores on all the pre-screened data. Finally, we re-train the TTS model with the filtered data.

% Section 4: Experimental evaluation
\vspace{-2mm}
\section{Experimental evaluation} \vspace{-1mm}
% 4.1. 実験条件
\subsection{Experimental conditions} %\vspace{-1mm}
\vspace{-1.5mm}
\subsubsection{Dataset} \vspace{-1.5mm}
%\textbf{Dark data.} 
We followed JTubeSpeech scripts~\cite{JTubeSpeech}
~\cite{JTubeSpeech_git}
%\footnote{\scriptsize\url{https://github.com/sarulab-speech/jtubespeech}} 
to obtain dark data from YouTube; the amount was approximately $3,500$ hours.
%\textbf{Pre-screened data.}
Pre-screening with a CTC threshold of $-0.3$ and speaker compactness threshold of $[1, 7]$\footnote{
    These values are the same as the experiments in the JTubeSpeech paper~\cite{JTubeSpeech}.
  },
  we obtained approximately $66$~hours ($60,000$~Japanese utterances) of $2,719$ speakers as the pre-screened data.
%\textbf{Data for pseudo MOS.}
The sentences used for calculating the pseudo MOS were $100$ phoneme-balanced sentences from the JVS corpus~\cite{JVS}. 
% \textbf{Test data.}
The test data used to evaluate the finally trained TTS models
   was $324$ sentences from the ITA corpus~\cite{ita}. 
% overlap
There was no overlap in text among the pre-screened data,
  sentences for the pseudo MOS, and test data.

\vspace{-3mm}
\subsubsection{Model and training} 
\label{sec:ModelAndTraining}
\vspace{-2mm}
% モデルと学習に関するものは全てここに書く
% \textbf{TTS model.}
We used FastSpeech~2~\cite{FastSpeech2} for our multi-speaker TTS model
  and the pre-trained HiFi-GAN vocoder~\cite{Hifi-gan} UNIVERSAL\_V1~\cite{hifigan_git}.
  %\footnote{\url{https://github.com/jik876/hifi-gan}}.
We followed the model size and hyperparameters of 
  the open-sourced implementation~\cite{FastSpeech2-JSUT}
  %\footnote{\scriptsize\url{https://github.com/Wataru-Nakata/FastSpeech2-JSUT}} 
  except for the speaker representation.
% $x$-vectorの話
Instead of the one-hot speaker representation implemented in the repository, 
  we used an open-sourced $x$-vector extractor~\cite{jtube_xvector},
  %\footnote{\scriptsize\url{https://github.com/sarulab-speech/xvector_jtubespeech}}, 
  and a $512$-dimensional $x$-vector was used to condition the TTS model. 
The $x$-vector was added to the output of the FastSpeech~2 encoder via a $512$-by-$256$ linear layer. 
The $x$-vector was averaged for each speaker; one $x$-vector corresponded to one unique speaker.
% 事前学習の話
The TTS model was pre-trained using $10,000$ utterances from the JVS corpus~\cite{JVS}, 
  the $100$-speaker Japanese TTS corpus. 
We performed 300k steps with a batch size of $16$ in this pre-training.
TTS training in this paper started from this pre-trained model
 with 100k steps with a batch size of $16$.

% \textbf{MOS prediction model.}
We used a pre-trained UTMOS~\cite{UTMOS} strong learner~\cite{UTMOS_git}
  %\footnote{\scriptsize\url{https://github.com/sarulab-speech/UTMOS22}}
  to obtain a five-scale pseudo MOS on naturalness from synthetic speech. 
%
% \textbf{Regression model to score training data.}
The regression model for predicting the pseudo MOS from the training data was 
  $1$-layer $256$-unit bi-directional long short-term memory~\cite{hochreiter1997long}, 
  followed by a linear layer, ReLU activation, and another linear layer. 
We used frame-level self-supervised learning~(SSL) features\footnote{
   We compared the SSL features and the NISQA~\cite{NISQA} features (used in the baseline) in the preliminary evaluation. The result demonstrated that the SSL features performed better.
  } 
  obtained with a wav2vec~2.0 model~\cite{wav2vec2}~\cite{wav2vec_git},
  %\footnote{\scriptsize\url{https://github.com/facebookresearch/fairseq/tree/main/examples/wav2vec}}, 
  as the input.
The frame-level outputs were aggregated to predict the pseudo MOS.
% 学習条件
The number of training steps, minibatch size, optimizer, and training objective were 10k, $12$, Adam~\cite{Adam} with a learning rate of $0.0001$, and mean squared error, respectively. 

\vspace{-3mm}
\subsubsection{Compared methods} 
\vspace{-2mm}
\label{sec:比較手法}
% 手法を itemize で説明．
We compared the following data selection methods.
\begin{itemize} \leftskip -5mm \itemsep -0mm
    \item \textbf{Unselected}: %  (no data selection)
    All the pre-screened data was used for the TTS training; 
      the training data size was approximately $60,000$ utterances. 
    \item \textbf{Acoustic-quality~(utterance-wise)}: 
    The training data was selected in terms of the acoustic quality of the data. 
    We used NISQA~\cite{NISQA}, a recent deep learning-based model to predict scores on 
      naturalness, noisiness, coloration, discontinuity, and loudness of the speech data. 
    Each score takes $[1,5]$, and we set the threshold to $3.5$, 
      i.e., data for which all the scores were higher than $3.5$ were selected. 
    The TTS training data size is approximately $12,000$. 
   \item \textbf{Ours-Utt~(evaluation-in-the-loop utterance-wise selection)}: 
   Our evaluation-in-the-loop data selection. 
   For each data, 
     we estimate the speech quality synthesized 
     by the TTS model from an initial training 
       with pre-screened data.
   The threshold for selecting training data was set to 
     have the resulting training data be the same in size as ``Acoustic-quality''.
   \item \textbf{Ours-Spk~(evaluation-in-the-loop speaker-wise selection)}:
   Our data selection, but the data selection was performed per speaker as described in Section~\ref{section:回帰モデルに関する説明}. 
   The threshold for selecting training data was set to 
     have the resulting the size of the training data be almost the same as ``Acoustic-quality''.
\end{itemize}

\vspace{-6mm}
\subsubsection{Evaluation} \vspace{-2mm}
\label{sec:実験結果の評価方法}
We evaluated the selection methods to clarify the following:
\vspace{-2mm}
\begin{itemize}\leftskip -5mm \itemsep 0mm
    \item \textbf{Does our method obtain more ``high-quality speakers?'': pseudo MOS comparison.} Our TTS model is expected to reproduce voices for a higher number of speakers. We define a ``high-quality speaker'' as a speaker with a higher pseudo MOS than the threshold. 
    % 閾値の決め方
    % JVSで学習した高品質な多話者TTSを学習
    % JVSの全員についてspeakersのpseudo MOSを計算
    % 最低値を閾値として設定
    We in advance trained an high-quality multi-speaker TTS model
      using the JVS corpus~\cite{JVS}
      and calculated the speaker-wise pseudo MOS scores.
    We set the lowest score among the JVS speakers as the threshold\footnote{
        The JVS corpus was constructed in a well-designed environment, 
          and we confirmed that it was not an outlier.
    }.
    Speakers with a higher score than the threshold were considered to be high-quality speakers. 
    % これを追記
    For each data selection method, we calculated
      1) the distribution of the pseudo MOSs for synthetic speech by the trained TTS model
        and 
      2) the number of high-quality speakers. 
    \item \textbf{Does our method work for unseen speaker?: pseudo MOS comparison.} The performance of the multi-speaker TTS model affects the synthetic speech quality of unseen speakers. Using the $x$-vector for unseen speakers, we counted the number of high-quality speakers among seen and unseen speakers, respectively.
    \item \textbf{Does our method increase speaker variation?} 
    % Therefore, we calculate cost of the minimum spanning tree of high-quality speakers' $x$-vectors. Here, edges exist between all vertices, and cost of an edge is $L_2$ distance between the vertices.
    % 
    We evaluated whether our method obtains diverse (i.e., sounding different) high-quality speakers. 
    To quantify the speaker variation, we calculated the cost of a Euclidean minimum spanning tree~\cite{Euclid最小全域木} of $x$-vectors of the high-quality speakers. 
    The calculation is similar to the g2g (median of the distances to the nearest $x$-vector)~\cite{g2g},
     but we used summation instead of a representative value~(i.e., median) 
     because our purpose is to evaluate how widely speakers spread.
    \item \textbf{Does synthetic speech of so-called ``high-quality speakers'' truly sound natural?: actual MOS evaluation.} We evaluated whether our selection based on pseudo MOS is truly effective in synthesizing perceptually high-quality speech. We subjectively evaluated the synthetic speech quality for the data selection methods and the performance, including the relationship 
    with pseudo MOS.
\end{itemize}

% ここで、psuedo MOS compasinにおいて各手法で話者が
Note that seen and unseen speakers are different among the data selection methods. 
All the speakers were seen ones for ``Unselected,'' but only parts of them were seen in the other methods. Also, seen speakers were different among ``Acoustic-quality,'' ``Ours-Utt,'' and ``Ours-Spk.'' 
Speakers not used for training 
  were considered to be unseen speakers for each method.
Unless otherwise noted, we describe results aggregating those of both seen and unseen speakers.

\vspace{-3.5mm}
\subsection{Results}
\vspace{-2mm}
\subsubsection{Number of high-quality speakers}
\label{sec:PseudoMOSExperimentResult}
\vspace{-2mm}

\begin{figure}[t]
  \centering
  \begin{minipage}{0.54\linewidth}
    \centering
    \includegraphics[width=1.\linewidth]{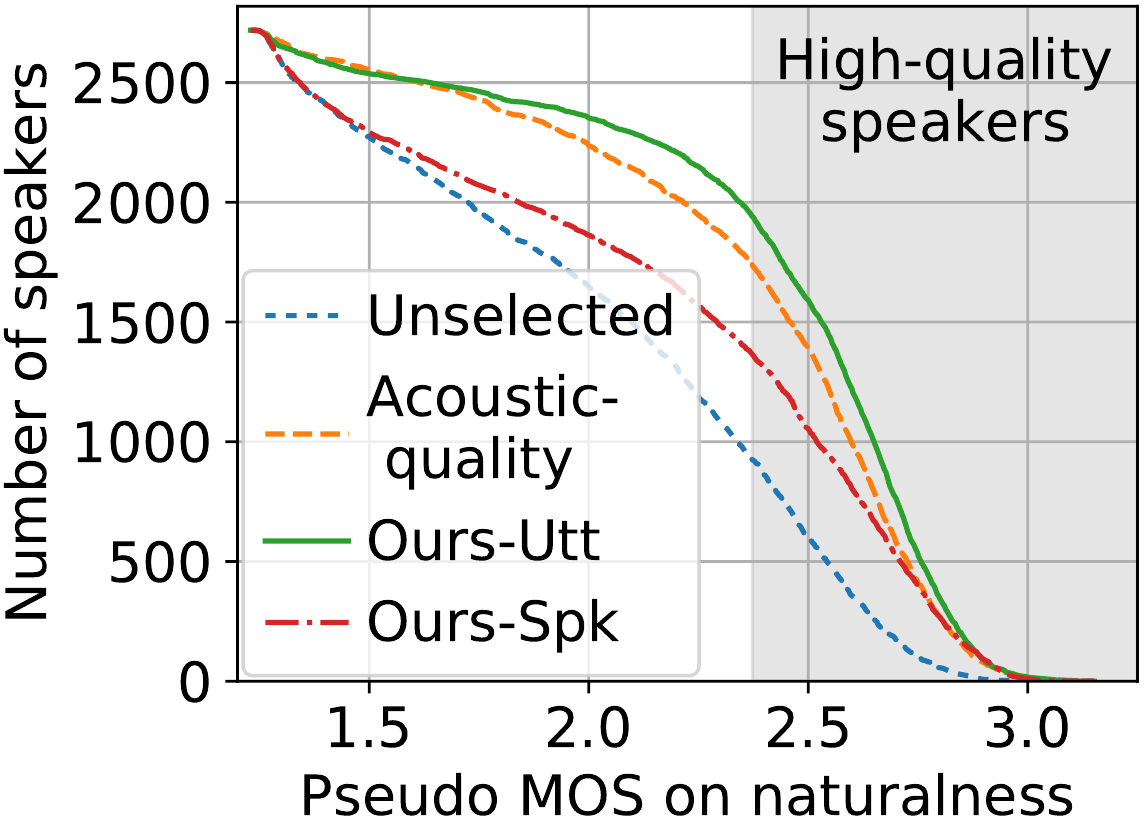}
    \vspace{-5mm}
    \subcaption{Pseudo MOS}
    \label{fig:MOS-話者数グラフ}
  \end{minipage}
  \hfill
  \begin{minipage}{0.45\linewidth}
    \centering
    \includegraphics[width=1.\linewidth]{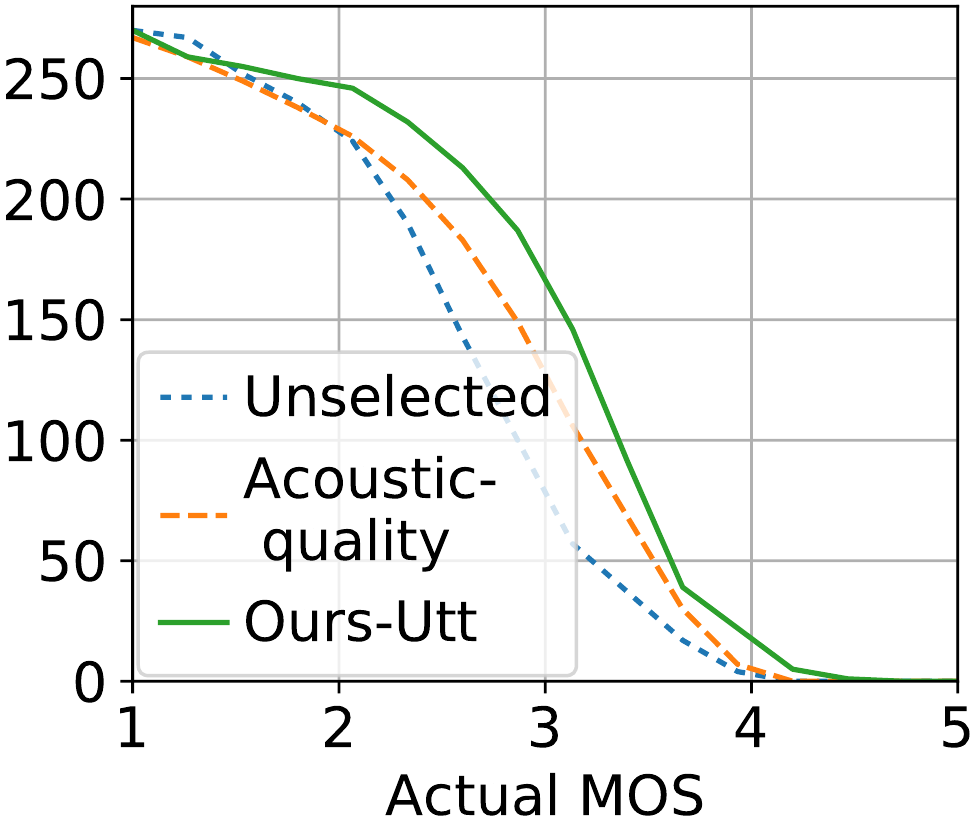}
    \vspace{-5mm}
    \subcaption{Actual MOS}
    \label{fig:主観評価実験グラフ}
  \end{minipage}
  \vspace{-3mm}
  \caption{Cumulative histograms of pseudo MOS and actual MOS. 
   Y-axis value indicates number of speakers 
     with higher score than x-axis value.}
  \vspace{-3mm}
  \label{fig:MOS-累積話者数分布グラフ}
\end{figure}

Fig.~\ref{fig:MOS-話者数グラフ} is a cumulative histogram of the pseudo MOSs. 
Our method had the highest values among the methods, 
  demonstrating that a TTS model trained on our data selection can synthesize
  multi-speaker voices with higher quality than the other methods. 
The numbers of high-quality speakers for 
   ``Unselected,'' ``Acoustic-quality,'' ``Ours-Utt,'' and ``Ours-Spk'' 
   were $924, 1737, 1942$, and $1367$, respectively.
We see that the proposed method increased the number of high-quality speakers, 
  compared with the other methods. 
Specifically, the increment was approximately $1.2$ times from that of ``Acoustic-quality.'' 
From these results, 
 we can say that the proposed method work better than the conventional methods
 for the purpose of increasing the speaker variety of the multi-speaker TTS model.
% 従来の日本語コーパスは100人規模のものであり今回のコーパスはそれらと比較して大規模なものである、という点は特筆に値する。
Note that this TTS corpus is much larger than the previous Japanese multi-TTS corpus~(JVS) composed of 100 speakers.

In addition, comparing ``Ours-Utt'' and ``Ours-Spk'',
  the proposed method was significantly better in terms of both
    the pseudo MOS distribution and the number of the high-quality speaker. 
This indicates that utterance-wise selection significantly contributes 
  to enhancing the performance, rather than speaker-wise selection.

\vspace{-3.5mm}
\subsubsection{Performance of unseen speakers}
\vspace{-2mm}

\begin{figure}[t]
  \centering
  \begin{minipage}{\linewidth}
    \centering
    \includegraphics[width=1.\linewidth]{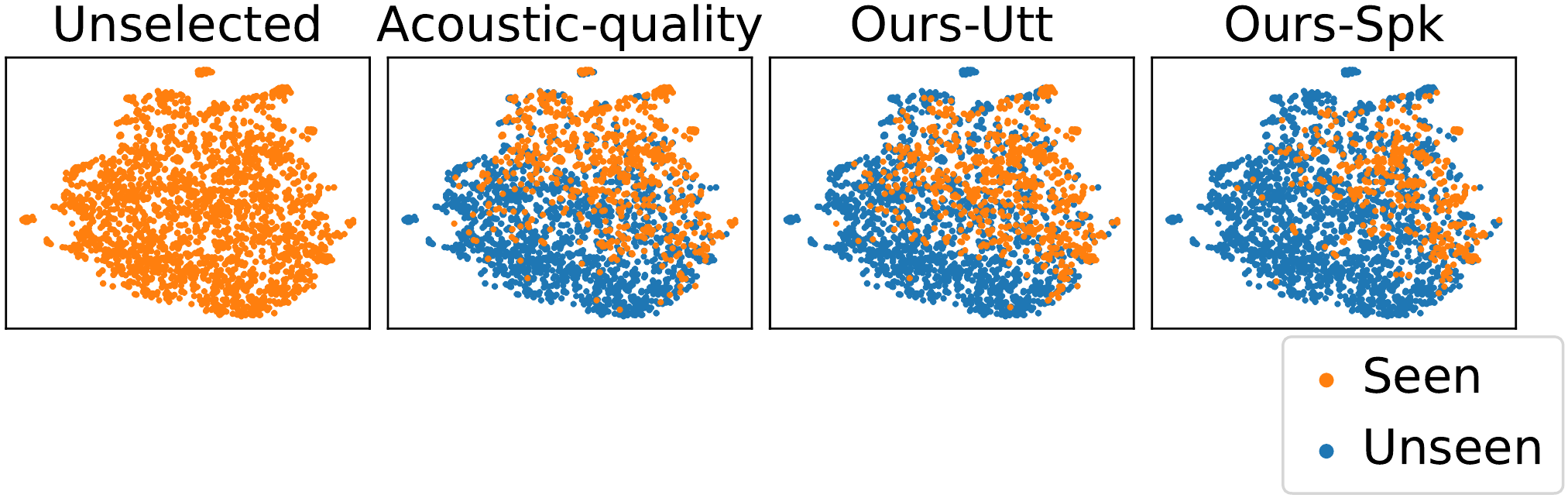}
    \vspace{-12mm}
    \subcaption{Seen/unseen speakers}
    \label{fig:seenとunseenの話者分布}
    \vspace{3.5mm}
  \end{minipage}
  \begin{minipage}{\linewidth}
    \centering
    \includegraphics[width=1.\linewidth]{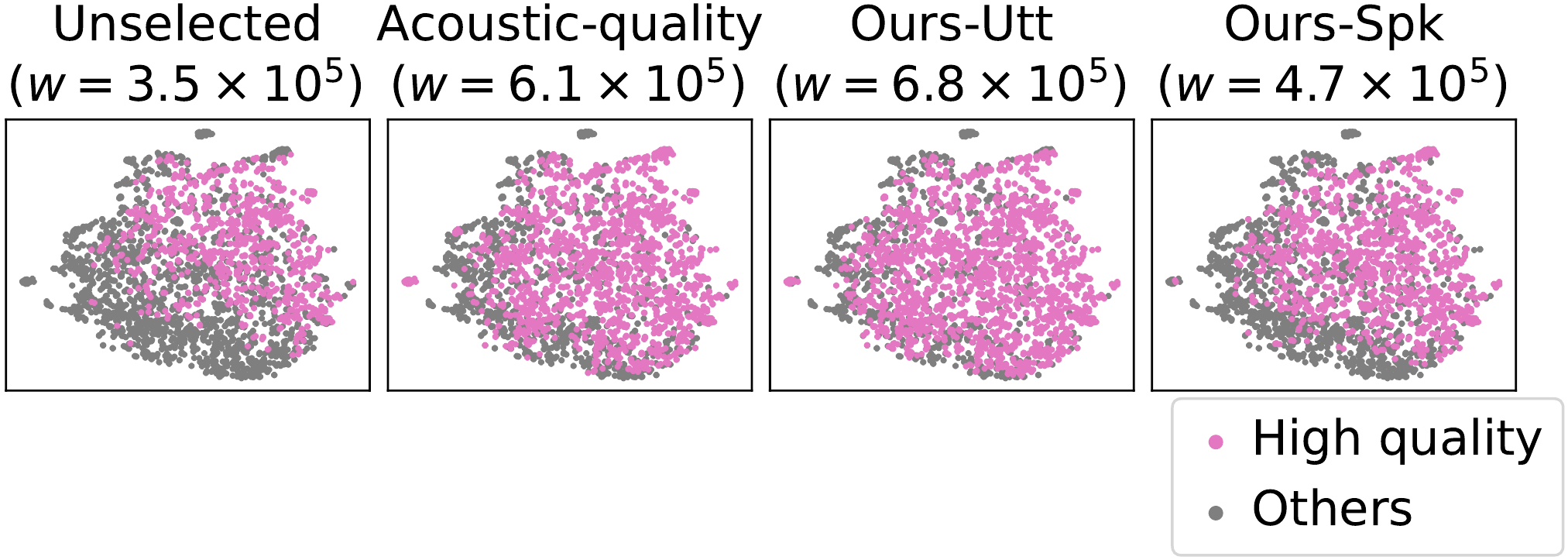}
    \vspace{-13mm}
    \subcaption{High-quality speakers.\\ 
        Higher $w$ indicates more diversity.}
    \label{fig:高品質話者の分布}
    \vspace{2mm}
  \end{minipage}
  \vspace{-3mm}
 \caption{ 
   Distributions of speakers by each data selection method.
  }
  \vspace{-3mm}
\end{figure}

\begin{comment}
\begin{figure}[t]
\centering
\includegraphics[width=0.9\linewidth]{image/seen_unseen.pdf}
 \vspace{-11mm}
 \caption{ 
   Distributions of seen/unseen speakers for\\
   each data selection method.
  }
 \vspace{-5mm}
 \label{fig:seenとunseenの話者分布}
\end{figure}
\end{comment}

\begin{table}[t]
  \caption{Number of seen and unseen speakers. Values of each cell are number of high-quality speakers, all speakers, and ratio of two values, respectively. 
  }
  \label{table:seen_unseen}
  \vspace{-2mm}
  \centering
  \footnotesize
  \begin{tabular}{r|c|c}
    Method  & Seen  & Unseen  \\ \hline
    Unselected & $ 924 / 2719 ( 34.0 \%) $ & - \\
    Acoustic-quality & $ 731 / 912 ( 80.2 \%) $ & $ 1006 / 1807 ( 55.7 \%) $ \\
    Ours-Utt & $ 811 / 882 ( 92.0 \%) $ & $ 1131 / 1837 ( 61.6 \%) $ \\
    Ours-Spk & $ 468 / 505 ( 92.7 \%) $ & $ 899 / 2214 ( 40.6 \%) $ \\
  \end{tabular}
  \vspace{-2mm}
\end{table}

Fig.~\ref{fig:seenとunseenの話者分布} shows the distributions of the seen/unseen speakers to qualitatively compare the data selection methods. 
Compared with ``Acoustic-quality,'' ``Ours-Utt'' had a similar variation.
% but filters out some speakers from the training data. 
Also, compared with ``Ours-Spk,'' ``Ours-Utt'' covered a wider range of speakers.  

Table~\ref{table:seen_unseen} lists the number of seen and unseen speakers for each data selection method. 
``Ours-Utt'' outperformed ``Acoustic-quality'' in terms of the percentage and absolute value for both seen and unseen speakers; our method worked even for unseen speakers.
``Ours-Spk'' was good at the percentage of seen speakers but not at the others. We expect that the small size of high-quality seen speakers in ``Ours-Spk'' caused this result.

\vspace{-3.5mm}
\subsubsection{Evaluation of speaker variation}
\vspace{-2mm}

%%%%%%%% この図は別の
\begin{comment}
\begin{figure}[t]
\centering
\includegraphics[width=0.9\linewidth]{image/answer_to_fig1.pdf}
 \caption{ 
   Distributions of high-quality speakers. \\
   $w$ is the speaker variation score. \\
   Higher $w$ indicates more diverse in high-quality speakers.
  }
 \vspace{-3mm}
\end{figure}
\end{comment}

Fig.~\ref{fig:高品質話者の分布} shows the distributions of the high-quality speakers and speaker variation scores $w$. 
Both qualitatively and quantitatively, ``Ours-Utt'' increased the speaker variation compared with the other methods, indicating that our method contributes to speaker variation.

\vspace{-3.5mm}
\subsubsection{Comparison of pseudo MOS and actual MOS}
\vspace{-2mm}

% 相関係数はハックしやすいので、生データを残す価値がある(紙面に余裕がなければ消す)
\begin{figure}[t]
\centering
\includegraphics[width=0.9\linewidth]{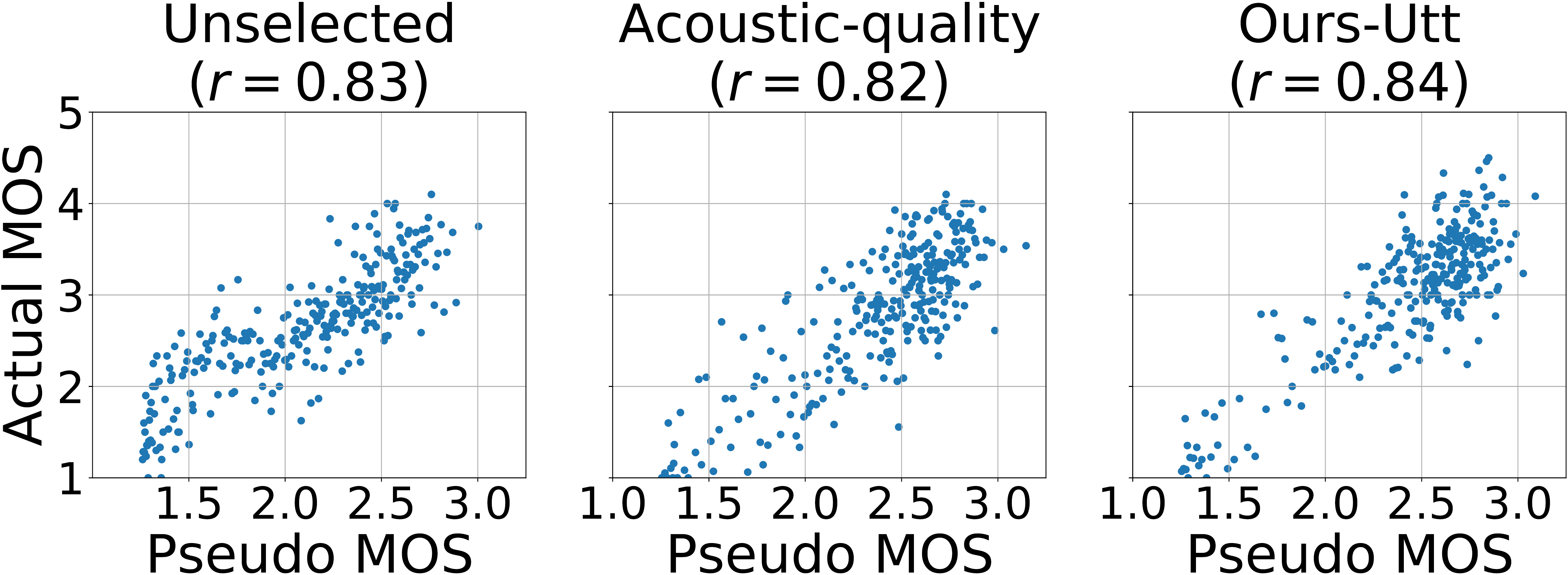}
 \caption{Pseudo MOS vs. actual MOS. Each point is each speaker.}
 \vspace{-3mm}
 \label{fig:疑似MOSvs実MOS}
\end{figure}

We conducted a five-point MOS test on the naturalness of synthetic speech. 
$500$ listeners participated, and each listener listened to $24$ samples. 
The TTS models except for ``Ours-Spk'' data selection were used for synthesizing speech. 
To reduce the evaluation cost, 
  we sampled speakers to be evaluated rather than using all the $2,719$ speakers. 
For each data selection method, 
  we divided the pseudo MOS in Fig.~\ref{fig:MOS-話者数グラフ} into $272$ intervals 
  and randomly selected one speaker from each interval. 
Therefore, $272$ speakers ($10$\% of $2,719$ speakers) were prepared for 
  each method. 
The actual MOSs were aggregated for each speaker. 

Fig.~\ref{fig:主観評価実験グラフ} shows the result. 
The proposed method had the highest values among all the methods. 
% この結果より，提案法が，知覚的な音質においても従来法より有効であることがわかる．
This result indicates that 
  the proposed method is more effective than the conventional methods
  even in perceptual speech quality.

% この結果について更に分析するために，疑似MOSと実MOSの関係を調査した．
To further analyze these results, 
  we investigated the relationship between the pseudo MOS and actual MOS.
% Fig.XX はその散布図，表XXは疑似MOSと実MOSの相関値である．
Fig.~\ref{fig:疑似MOSvs実MOS} shows scatter plots 
  and correlation coefficients $r$.
% これらの結果より，疑似MOSと実MOSの相関は非常に高く，全てのデータ選択法においてその値が0.8を超えている．
% 本実験で使用された疑似MOS予測モデルは英語と中国語の合成音声で学習されており，本実験の対象である日本語は学習に含まれていない．
These results show that 
    the correlation was always high~($r>0.8$)
  despite our target language~(Japanese)  
    not being included in the training data~(English and Chinese) of 
    the pseudo MOS prediction model.
%   1) 話者選択に疑似MOSの利用が有効であり，
%   2) 疑似MOSの相対的な大小\footnote{
%     なお，Section 2.Xで言及したように，疑似MOS値そのものは，
%     実MOSのそれと大きく異なることに注意する．
%   }は，学習データに含まれない言語においても十分に有効であることを表す．
This indicates that
  pseudo MOS is valid for languages
  not included in pseudo MOS training data,
  indicating a good capability for languages.

\vspace{-4mm}

% Section 5: Conclusion
\section{conclusion}
\vspace{-2.2mm}
We proposed an evaluation-in-the-loop data selection method for TTS from dark data. Experimental results using YouTube videos showed that our method significantly outperformed the conventional acoustic-quality-based method. 
Our future work includes the use of more recent TTS models. 

% bib
\newpage
%{\footnotesize
% \bibliographystyle{bib/IEEEbib}
\printbibliography

\end{document}